\documentclass[preprint,amsmath,amssymb,prl]{revtex4}
\usepackage{graphicx}
\usepackage{bm}
\usepackage{subfigure}

\begin{document}
\title{Strongly quadrature-dependent noise in superconducting micro-resonators measured at the vacuum-noise limit}

\author{J. Gao}
\affiliation{National Institute of Standards and Technology,
Boulder, CO 80305-3337}

\author{L. R. Vale}
\affiliation{National Institute of Standards and Technology,
Boulder, CO 80305-3337}

\author{J. A. B. Mates}
\affiliation{National Institute of Standards and Technology,
Boulder, CO 80305-3337}

\author{D. R. Schmidt}
\affiliation{National Institute of Standards and Technology,
Boulder, CO 80305-3337}

\author{G. C. Hilton}
\affiliation{National Institute of Standards and Technology,
Boulder, CO 80305-3337}

\author{K. D. Irwin}
\affiliation{National Institute of Standards and Technology,
Boulder, CO 80305-3337}

\author{F. Mallet}
\affiliation{JILA and the Department of Physics, National Institute
of Standards and Technology and University of Colorado, Boulder, CO
80309-0440}

\author{M. A. Castellanos-Beltran}
\affiliation{JILA and the Department of Physics, National Institute
of Standards and Technology and University of Colorado, Boulder, CO
80309-0440}

\author{K. W. Lehnert}
\affiliation{JILA and the Department of Physics, National Institute
of Standards and Technology and University of Colorado, Boulder, CO
80309-0440}

\author{J. Zmuidzinas}
\affiliation{Jet Propulsion Laboratory, California Institute of
Technology, Pasadena, CA 91125}

\author{H. G. Leduc}
\affiliation{Jet Propulsion Laboratory, California Institute of
Technology, Pasadena, CA 91125}

\begin{abstract}
We measure frequency- and dissipation-quadrature noise in
superconducting lithographed microwave resonators with sensitivity
near the vacuum noise level using a Josephson parametric amplifier.
At an excitation power of 100~nW, these resonators show significant
frequency noise caused by two-level systems. No excess
dissipation-quadrature noise (above the vacuum noise) is observed to
our measurement sensitivity.
These measurements demonstrate that the excess
dissipation-quadrature noise is negligible compared to vacuum
fluctuations, at typical readout powers used in micro-resonator
applications. Our results have important implications for resonant
readout of various devices such as detectors, qubits and
nano-mechanical oscillators.
\end{abstract}

\date{\today}

\maketitle

\begin{figure}[h]
          \includegraphics[width=\linewidth]{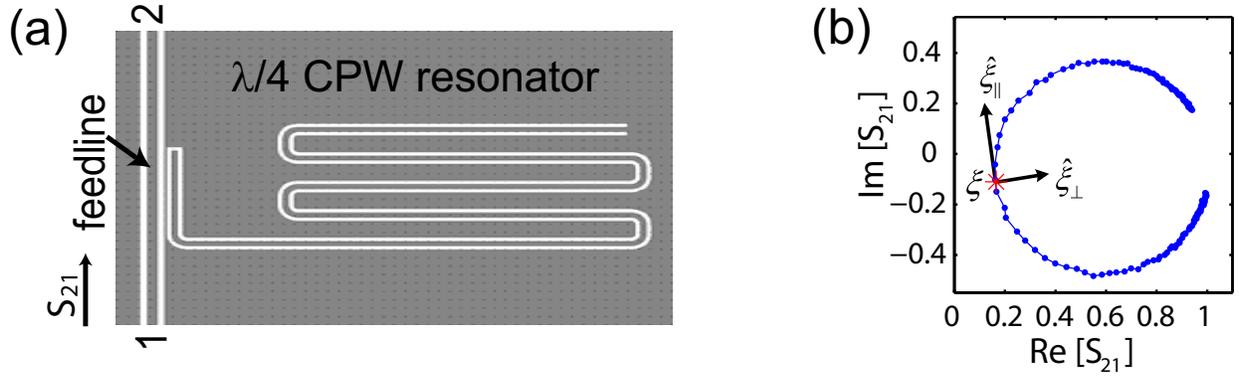}
          \caption{(a) Schematic illustration (not to scale) of the resonator
and feedline (Nb in gray and Si in white).
          (b) Transmission $S_{21}$ measured with a vector network analyzer (VNA). A parametric plot of Im$[\mathrm{S}_{21}(f)]$ vs Re$[\mathrm{S}_{21}(f)]$ traces out a resonant circle. The resonance point, frequency
quadrature (tangential to the circle) and dissipation quadrature
(normal to the circle) are indicated by $\xi$,
$\hat{\xi_{\parallel}}$ and $\hat{\xi_{\perp}}$, respectively.
}\label{fig:ress21}
\end{figure}

Superconducting lithographed microwave resonators (micro-resonators)
are used in a broad range of applications,
such as photon detection\cite{Day03}, readout of superconducting
quantum interference devices (SQUIDs)\cite{Mates08}, measurement of
qubits\cite{Wallraff04} and detection of nanomechanical
motion\cite{Regal2008}. In these applications, the sensing element
presents a variable reactive or dissipative load to the
micro-resonator. The microwave probe signal, transmitted through or
reflected from the resonator, is sensitive to changes in resonance
frequency $f_\mathrm{r}$ in one quadrature of the microwave field
(referred to as the frequency quadrature, $\hat{\xi_{\parallel}}$ in
Fig.~\ref{fig:ress21}), and changes in quality factor $Q_\mathrm{r}$
(or internal dissipation) in the orthogonal quadrature (referred to
as the dissipation quadrature, $\hat{\xi_{\perp}}$ in
Fig.~\ref{fig:ress21}). The use of these two quadratures is often
referred to as frequency readout and dissipation readout,
respectively.

The sensitivity of these measurements may be limited by the noise in
the microwave probe signal, the intrinsic noise of the resonator,
and noise in the readout electronics, usually set by the noise
temperature $T_\mathrm{n}$ of the amplifier. These noise
contributions can also be projected into the two quadratures, with
power spectra $S_{\parallel}$ and $S_{\perp}$ for
$\hat{\xi_{\parallel}}$ and $\hat{\xi_{\perp}}$, respectively. In
general, the noise power can be quadrature-dependent; $S_{\parallel}
\neq S_{\perp}$. If both the resonator and the amplifier have no
excess noise, the ultimate sensitivity will be limited quantum
mechanically by the vacuum noise, which is a quarter photon per
quadrature ($S_{\parallel}=S_{\perp}=hf/4$).

We have previously reported that at high excitation power
significant excess noise in the frequency quadrature is universally
observed in superconducting micro-resonators
 and suggested that a surface layer of two-level system (TLS) fluctuators are responsible
 for this noise\cite{Gao2007, Gao2008b}. In contrast, no excess noise in the
dissipation quadrature was observed above the noise floor of a high
electron mobility transistor (HEMT) amplifier. However, a
state-of-the-art HEMT has $T_\mathrm{n}\approx$~5 K at 6 GHz, which
is 35 times the vacuum noise. Therefore, it has been unclear whether
micro-resonators produce excess noise in the dissipation quadrature
above the vacuum noise level. On the other hand, microwave
amplifiers based on SQUIDs, including dc-SQUID
amplifiers\cite{Michael2003, Spietz2009} and Josephson parametric
amplifiers (JPA)\cite{Yurke1989}, have recently demonstrated nearly
ideal performance. For example, our group has demonstrated a JPA
that is tunable between 4-8 GHz and adds only 0.3 photon of
noise\cite{Castellanos-Beltran2008}. It is possible, and indeed
promising to use these quantum amplifiers to improve the sensitivity
of resonator dissipation measurements, provided that the excess
noise in the dissipation quadrature of the resonator is well below
the HEMT noise floor.

In this Letter, we report the measurement of the resonator noise
using a Josephson parametric amplifier. We have achieved a system
noise floor as low as 1.8 times the vacuum noise or a factor of 23
lower than a HEMT by using the JPA as a nearly noiseless
preamplifier. These measurements, for the first time, show that the
excess resonator noise in the dissipation quadrature can be below
the vacuum noise, at an excitation power that is appropriate for
kinetic inductance detector(MKID)\cite{Day03} readout.

We measured the noise of a quarter-wave coplanar waveguide (CPW)
resonator, made by patterning a 200~nm-thick Nb film deposited on a
high-purity intrinsic silicon
($15~\mathrm{k}\Omega\cdot\mathrm{cm}$) substrate. The CPW resonator
has a center strip width of $5~\mu$m, gaps of $1~\mu$m, and is
capacitively coupled to a CPW feedline (see
Fig.~\ref{fig:ress21}(a)). From the transmission $S_{21}$ data, we
have determined $f_\mathrm{r} = 6.23~$GHz and $Q_\mathrm{r}
=1.5\times 10^5$ (internal $Q_\mathrm{i}=9\times 10^5$).

\begin{figure}[h]
          \includegraphics[width=\linewidth]{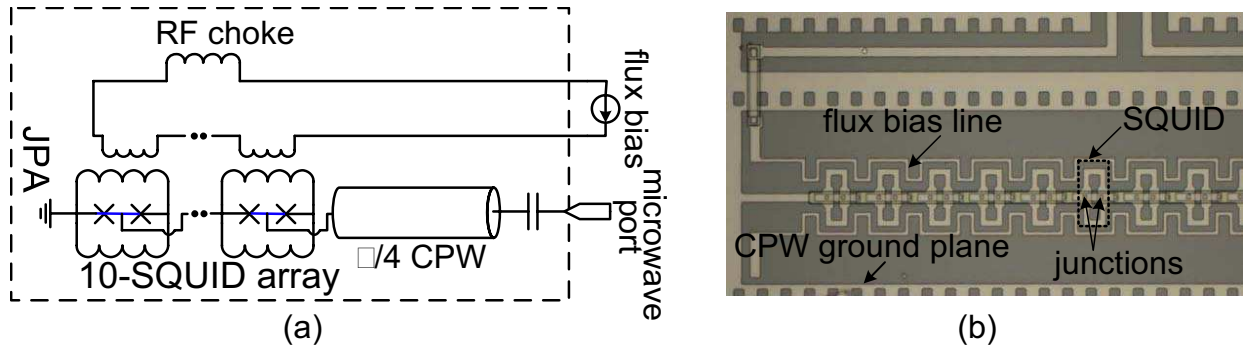}
         \caption{(a) Schematic of the JPA consisting an array of 10 parallel
         gradiometric SQUIDs coupled to a flux-bias line.
         (b) Microscope picture showing the SQUID array.}\label{fig:JPA}
\end{figure}

Another crucial component for this experiment is a Josephson
parametric amplifier, which uses the nonlinear inductance of a
Josephson junction to achieve parametric
amplification\cite{Yurke1989}. The JPA used in this experiment
consists of a Nb on Si quarter-wave CPW resonator whose center
conductor is terminated by an array of 10 SQUIDs in series. A
schematic of the JPA and a microscope picture of the SQUIDs are
shown in Fig.~\ref{fig:JPA}. Here we use a parallel gradiometric
SQUID design to minimize the magnetic noise pick-up. A flux-bias
line with a built-in RF-choke is used to tune the center frequency
of the JPA.

\begin{figure}[h]
          \includegraphics[width=\linewidth]{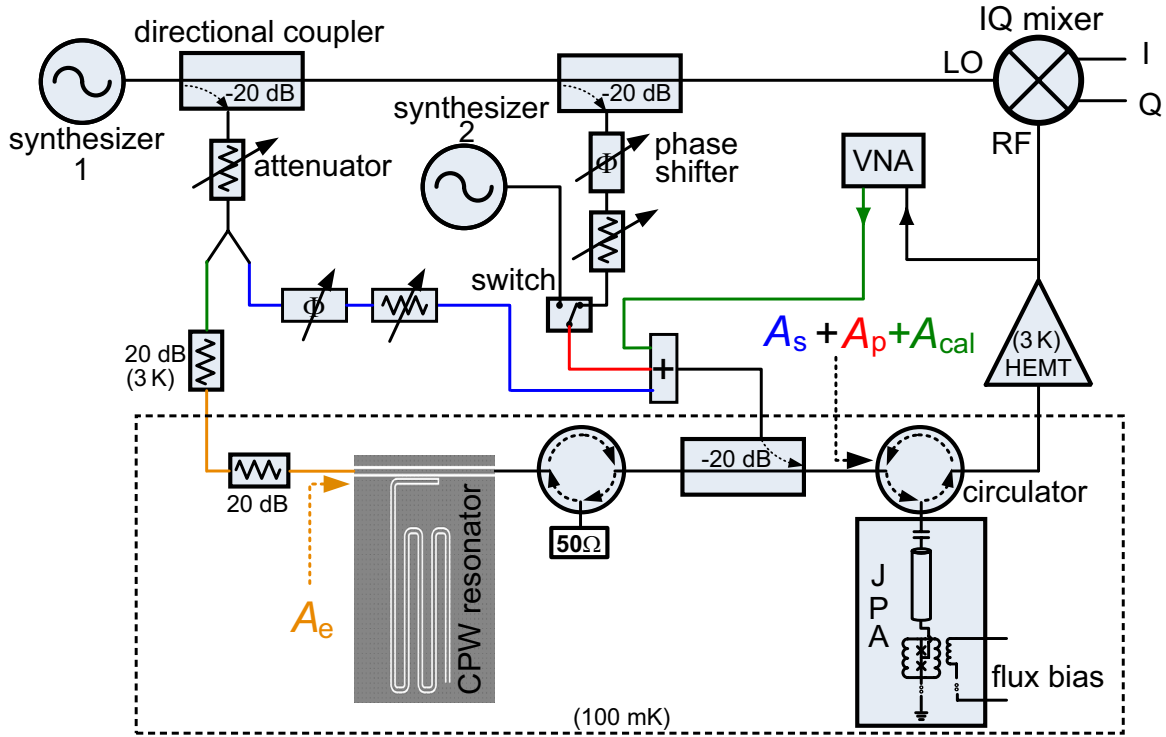}
          \caption{A diagram of noise measurement setup. The calibration tone $A_\mathrm{cal}$
          is generated by the VNA that is also used for $S_{21}$ measurement.
          A switch is used to select the source of the pump tone between synthesizer 1 and synthesizer 2,
          which allows the JPA to operate in degenerate mode (switch to the right) or nondegenerate mode (switch to the left).}
\label{fig:JPARESfull}
\end{figure}

We connect the resonator to the JPA in a configuration shown in
Fig.~\ref{fig:JPARESfull}. Four microwave signals with phase and/or
amplitude control are generated: a resonator excitation tone
$A_\mathrm{e}$ to excite the resonator, a carrier suppression tone
$A_\mathrm{s}$ to cancel the coherent carrier ($A_\mathrm{e}
S_{21}$) in the transmitted signal, a pump tone $A_\mathrm{p}$ to
pump the JPA, and a calibration tone $A_\textrm{cal}$ that is always
detuned from $A_e$ by 100~kHz, to calibrate the noise measurement.
The three tones $A_\mathrm{s}$, $A_\mathrm{p}$ and $A_\textrm{cal}$
are injected through a directional coupler between the resonator and
the JPA. After carrier suppression, only the resonator noise signal
is sent to the JPA for amplification, which prevents the JPA from
saturating and also allows the phase of $A_\mathrm{p}$ to be
independently adjusted relative to the noise signal. The resonator
and JPA, along with other components inside the dashed rectangle in
Fig.~\ref{fig:JPARESfull}, are cooled in an adiabatic
demagnetization refrigerator (ADR) to 100~mK.

We first characterize our JPA by turning off the $A_\mathrm{e}$ and
$A_\mathrm{s}$ tones. When $A_\mathrm{p}$ is also turned off and no
flux bias is applied, the JPA shows a resonance frequency of
$f_\mathrm{JPA} = 6.80~$GHz and $Q_\mathrm{JPA}\sim150$
(coupling-limited). With flux bias, $f_\mathrm{JPA} $ is tunable
between 5.8~GHz and 6.80~GHz while $Q_\mathrm{JPA}$ remains
constant. When $A_\mathrm{p}$ is turned on, the JPA gives a power
gain $0 \leq G \leq 35$~dB and an amplitude gain-bandwidth product
of 45~MHz\cite{Castellanos-Beltran2008}. The highest gain occurs at
a critical pump power $P_\mathrm{c} \approx -93$~dBm (500~fW). Above
$P_\mathrm{c}$, the JPA response is bistable\cite{Siddiqi2004}.

\begin{figure}[h]
          \includegraphics[width=0.85\linewidth]{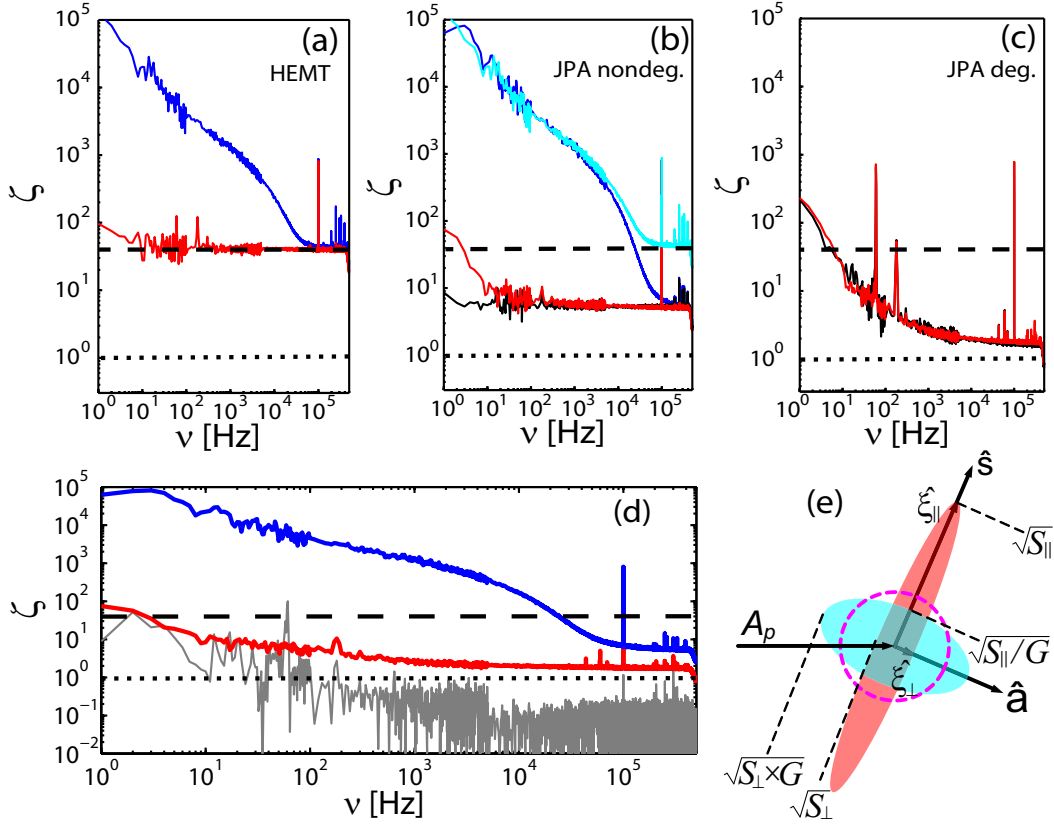}
         \caption{Measured noise spectra $S_{\parallel}$ (blue) and $S_{\perp}$ (red), directly by HEMT (a), by JPA in nondegenerate mode (b) and JPA in degenerate mode ($S_{\perp}$ only)(c).
         The HEMT noise floor (dashed line), the system noise floor--measured with $A_\mathrm{e}$ and $A_\mathrm{s}$ off--(solid black line), and the vacuum noise level (dotted line) are indicated in each subfigure.
         The lowest measured $S_{\perp}$ (the minimum of the red lines in (a), (b), and (c)) is plotted by the red line in (d), which, after subtracting off the system noise floor (including HEMT, thermal and vacuum noise), yields the gray
         line. All the noise spectra are plotted in units of vacuum
         noise and are calculated by
         $\zeta= S_\mathrm{\xi}/(hf/4)$, where
         $S_\mathrm{\xi}$ is the quadrature noise power spectral density(double-sided) in units of
         $\mathrm{W}$/Hz measured at the output of IQ mixer expressed as apparent noise at the input of
         JPA. A calibration tone of known power, $P_\mathrm{cal} = -153.2$~dBm, is injected at the
         JPA input, allowing us to infer the gain of the
         amplification and mixing chain. $P_\mathrm{cal}$ is known, because the critical pump power $P_\mathrm{c}$ is accurately determined by
         a separate measurement performed in a liquid-helium test
         probe. In contrast to the ADR, for the test probe we are able
         to accurately calibrate the attenuation of the microwave
         cables carrying signals to the JPA\cite{cal}.
        The alignment of the resonator noise ellipse with the JPA
amplification and squeezing quadratures are illustrated in (e). The
input and output noise ellipses are shown in red and cyan,
respectively. The HEMT noise is indicated by the pink dashed circle.
         }\label{fig:noiseall}
\end{figure}

We then turn off $A_\mathrm{p}$ and detune the JPA frequency
$f_\mathrm{JPA}$ far from the resonator's frequency $f_\mathrm{r}$.
This allows us to measure the resonator noise directly with the
HEMT. The noise spectra $S_{\parallel}$ and $S_{\perp}$ measured at
an excitation (internal) power\cite{Gao2007} $P_\mathrm{int}\approx
-40$~dBm (100~nW) are plotted in Fig.~\ref{fig:noiseall}(a) in units
of vacuum noise $\zeta$ ($\zeta =1$ corresponds to vacuum noise
level). In the frequency quadrature, $S_{\parallel}$ rises well
above the HEMT noise floor and the level is consistent with that
reported in Ref. \cite{Gao2007}. In the dissipation quadrature, no
excess noise is seen above the HEMT noise floor. This HEMT noise
referred to the input port of the JPA is found to be $\zeta = 40$ or
$T_\mathrm{n} = 6$~K.

To reveal the actual resonator noise in the dissipation quadrature,
we use the JPA as a preamplifer before the HEMT. We first operate
the JPA in the ``nondegenerate'' mode by detuning $A_\mathrm{p}$
(generated by a second synthesizer) from both $A_\mathrm{e}$ and
$A_\mathrm{s}$ by 2.3~MHz. In this mode, the JPA is a linear
phase-insensitive amplifier that amplifies both quadratures with a
power gain chosen to be $G=16$. The results are shown in
Fig.~\ref{fig:noiseall}(b). In the frequency quadrature,
$S_{\parallel}$ matches that directly measured by the HEMT (the cyan
line). In the dissipation quadrature, except for a $1/f$ noise below
10~Hz, no excess noise is seen above the system noise floor (the
black line). With $G=16$, the JPA achieves a system noise of $\zeta
= 5.1$, a factor of $8$ better than the HEMT (dashed line). Out of
the $\zeta = 5.1$ system noise, the HEMT contributes $\zeta = 2.5$
(40/16), vacuum and thermal noise contributes $\zeta = 1.3$, and our
JPA adds $\zeta = 1.3$. According to the quantum theory of
amplifiers\cite{Caves1982}, a linear phase-insensitive amplifier
must add noise at least as large as the vacuum ($\zeta = 1$), which
is called the standard quantum limit (SQL). We infer that the JPA
itself has in fact added only $\zeta = 0.3$ noise above the SQL. By
operating at higher gain this JPA should be able to achieve a system
noise as low as $\zeta = 1.3$\cite{gain}.

Next, we measure the resonator noise with the JPA operating in the
``degenerate'' mode. In this mode, all the tones are generated by
the same synthesizer at the same frequency, which minimizes the
effect of synthesizer phase noise. The JPA in the degenerate mode is
a phase-sensitive amplifier that amplifies a specific quadrature
$\hat{a}$ of the input signal by a factor of $\sqrt{G}$ and squeezes
(deamplifies) its orthogonal quadrature $\hat{s}$ by the same
factor. In this measurement, we carefully align the resonator
dissipation quadrature $\hat{\xi_\perp}$ to the JPA amplification
quadrature $\hat{a}$ (see Fig.~\ref{fig:noiseall}(e)). With adequate
gain ($G \gtrsim 40$), the vacuum noise or any excess noise in the
dissipation quadrature is amplified and visible above the HEMT noise
floor (dashed pink circle). A power gain of $G \sim 100$ is used. To
accurately align $\hat{\xi_\perp}$ with $\hat{a}$, we adjust the
phase of $A_\mathrm{p}$ with respect to the input resonator noise
signal. When $\hat{\xi_\perp}$ and $\hat{a}$ are misaligned by an
angle $\theta$, the output spectrum for the amplified quadrature
$\hat{a}$ is $S^\mathrm{out}_\mathrm{aa} = G(S_\perp\cos^2\theta
+S_\parallel\sin^2\theta)$.
Because we know $S_\parallel \gg S_\perp$,
$S^\mathrm{out}_\mathrm{aa}/G \geq S_\perp$ and the minimum occurs
when $\hat{\xi_\perp}$ and $\hat{a}$ are perfectly aligned
($\theta=0$). The red line in Fig.~\ref{fig:noiseall}(c) shows the
lowest measured noise spectrum from the amplified quadrature of the
JPA referred to noise in the resonators. Again, we see no excess
noise above the system noise floor (black line). The system noise
(white noise above 10~kHz) corresponds to $\zeta = 1.8$ (a factor of
$23$ lower than the HEMT), with the HEMT accounting for $\zeta =
0.4$ (40/100), vacuum and thermal noise accounting for $\zeta =
1.3$, and the JPA adding $\zeta = 0.1$. According to the quantum
theory of amplifiers, there is no SQL for the noise added by an
amplifier that amplifies only one quadrature. Indeed, for modulation
frequencies about 10 kHz, our JPA is almost noiseless when operating
at high gain in the degenerate mode. However, we do see in
Fig.~\ref{fig:noiseall}(c) substantial noise below 10~kHz. We find
that the noise level and shape changes when $f_\mathrm{JPA}$ is
tuned away from $f_\mathrm{r}$. We suspect that the pump tone can
excite the resonator and add noise even when $A_\mathrm{e}$ is off,
due to the finite directivity (-40~dB total) of the circulators.

The results of the noise measurements made using the HEMT, and using
the JPA in the nondegenerate and degenerate modes, are combined in
Fig.~\ref{fig:noiseall}(d), where the red line plots the lowest
measured $S_{\perp}$ at each frequency. The excess resonator noise
in the dissipation quadrature is constrained by this noise line.
Because this noise line contains other noise contributions (HEMT,
vacuum and thermal noise), we can further tighten this constraint by
subtracting the system noise floor from the measured noise level.
This yields the gray line in Fig.~\ref{fig:noiseall}(d), which
reaches below the vacuum noise (the dotted line) at $\nu>100~$Hz.
Therefore, the excess resonator noise in the dissipation quadrature
is constrained to lie below the vacuum noise at modulation
frequencies above 100~Hz, which is the main conclusion of this
letter.

The property that the resonator noise is entirely directed in the
frequency quadrature, with $S_{\parallel}$ rising orders of
magnitude above and $S_{\perp}$ remaining below the vacuum noise, is
an extremely interesting observation. Previous studies, including
the power\cite{Gao2007}, temperature\cite{Kumar2008}, and geometry
dependence\cite{Gao2008b} of the frequency noise, its origin in the
resonator capacitance\cite{Omid2009}, and the anomalous frequency
shift below 1~K\cite{Gao2008a}, have led to a consistent picture in
which the noise is caused by a surface layer of TLS fluctuators in
the dielectric materials. In this picture, TLS saturation plays an
important role. Indeed, the average response
$\chi(\omega)$\cite{Gao2008b} of a TLS to the microwave field
$|\vec{E}|$ is given by\cite{Phillips1972}
\begin{equation}
\chi(\omega) \propto
\frac{\omega_\mathrm{0}-\omega}{(\omega_\mathrm{0}-\omega)^2+({s} /
T_2)^2} + j \frac{T_2^{-1}}{(\omega_\mathrm{0}-\omega)^2+({s} /
T_2)^2},\nonumber
 \label{eqn:ch1ch2}
\end{equation}
where $\omega_\mathrm{0}$ is the transition frequency of the TLS,
$T_2$ is the transverse relaxation time, ${s} =
[1+(|\vec{E}|/E_c)^2]^{-1/2}$ is the power broadening factor ($s
\gtrsim 100$ is estimated for our case) and $E_c$ is a critical
electric field\cite{Martinis2005}. While power-broadening expands
the line width by a factor of ${s}$, it also reduces the peak
strength of the reactive response by ${s}$, and the dissipative
response by $s^2$. Therefore the effect on the noise goes as
${s}^{-1}$ and ${s}^{-3}$, respectively in the two quadratures. This
may qualitatively explains the observed $P^{-1/2}$ ($\propto
{s}^{-1}$) power dependence of $S_{\parallel}$ and the large ratio
($\propto {s}^{2}$) between $S_{\parallel}$ and $S_{\perp}$.
However, although there is some speculation about the TLS noise
mechanism\cite{Gao2008b,Yu2004} and attempts to model this noise
\cite{Gao2008b, Constantin2009}, so far the physical picture of TLS
resonator noise is not well understood and a quantitative model is
not available.

In conclusion, we have demonstrated the measurement of resonator
noise using a Josephson parametric amplifier. We have achieved a
system noise that is 5.1 and 1.8 times the vacuum noise when
operating the JPA in nondegenerate and degenerate mode, which is a
factor 8 and 23 lower than the HEMT noise. At an excitation power of
100~nW, no excess noise in the dissipation quadrature is observed
above the system noise floor. Excess resonator noise in the
dissipation quadrature is constrained to be below the vacuum noise
above 100~Hz. We suggest that the frequency-only noise property is
related to the saturation of two-level systems.

Our results have important implications for resonant readout of
sensitive devices. If the sensing element presents a fully or
partially resistive load to the resonant circuit and thereby
produces a signal in the dissipation quadrature, the use of an
ultra-low-noise amplifier (well below the HEMT noise) in a
dissipation readout may improve the measurement sensitivity by a
large factor. For example, the resonator parameters and excitation
power used in this experiment are typical for MKID readout;
therefore the sensitivity of MKIDs can be readily improved by a
factor of $\sqrt{23}$ by using the JPA in this letter in degenerate
mode, without requiring modification to the detector design or
operating conditions.

We thank Jose Aumentado, Lafe Spietz, John Teufel and John Martinis
for useful discussions. The resonator was fabricated in the JPL
Microdevices Lab. The JPA was fabricated in the NIST, Boulder
fabrication facility. This work was supported in part by the DARPA
QUEST program.

\pagebreak

\begin{thebibliography}{21}
\expandafter\ifx\csname
natexlab\endcsname\relax\def\natexlab#1{#1}\fi
\expandafter\ifx\csname bibnamefont\endcsname\relax
  \def\bibnamefont#1{#1}\fi
\expandafter\ifx\csname bibfnamefont\endcsname\relax
  \def\bibfnamefont#1{#1}\fi
\expandafter\ifx\csname citenamefont\endcsname\relax
  \def\citenamefont#1{#1}\fi
\expandafter\ifx\csname url\endcsname\relax
  \def\url#1{\texttt{#1}}\fi
\expandafter\ifx\csname urlprefix\endcsname\relax\def\urlprefix{URL
}\fi \providecommand{\bibinfo}[2]{#2}
\providecommand{\eprint}[2][]{\url{#2}}

\bibitem[{\citenamefont{Day et~al.}(2003)\citenamefont{Day, LeDuc, Mazin,
  Vayonakis, and Zmuidzinas}}]{Day03}
\bibinfo{author}{\bibfnamefont{P.~K.} \bibnamefont{Day}},
  \bibinfo{author}{\bibfnamefont{H.~G.} \bibnamefont{LeDuc}},
  \bibinfo{author}{\bibfnamefont{B.~A.} \bibnamefont{Mazin}},
  \bibinfo{author}{\bibfnamefont{A.}~\bibnamefont{Vayonakis}},
  \bibnamefont{and}
  \bibinfo{author}{\bibfnamefont{J.}~\bibnamefont{Zmuidzinas}},
  \bibinfo{journal}{Nature} \textbf{\bibinfo{volume}{425}},
  \bibinfo{pages}{817} (\bibinfo{year}{2003}).

\bibitem[{\citenamefont{Mates et~al.}(2008)\citenamefont{Mates, Hilton, Irwin,
  Vale, and Lehnert}}]{Mates08}
\bibinfo{author}{\bibfnamefont{J.~A.~B.} \bibnamefont{Mates}},
  \bibinfo{author}{\bibfnamefont{G.~C.} \bibnamefont{Hilton}},
  \bibinfo{author}{\bibfnamefont{K.~D.} \bibnamefont{Irwin}},
  \bibinfo{author}{\bibfnamefont{L.~R.} \bibnamefont{Vale}}, \bibnamefont{and}
  \bibinfo{author}{\bibfnamefont{K.~W.} \bibnamefont{Lehnert}},
  \bibinfo{journal}{Appl. Phys. Lett.} \textbf{\bibinfo{volume}{92}},
  \bibinfo{eid}{023514} (\bibinfo{year}{2008}).

\bibitem[{\citenamefont{Wallraff et~al.}(2004)\citenamefont{Wallraff, Schuster,
  Blais, Frunzio, Huang, Majer, Kumar, Girvin, and Schoelkopf}}]{Wallraff04}
\bibinfo{author}{\bibfnamefont{A.}~\bibnamefont{Wallraff}},
  \bibinfo{author}{\bibfnamefont{D.~I.} \bibnamefont{Schuster}},
  \bibinfo{author}{\bibfnamefont{A.}~\bibnamefont{Blais}},
  \bibinfo{author}{\bibfnamefont{L.}~\bibnamefont{Frunzio}},
  \bibinfo{author}{\bibfnamefont{R.-S.} \bibnamefont{Huang}},
  \bibinfo{author}{\bibfnamefont{J.}~\bibnamefont{Majer}},
  \bibinfo{author}{\bibfnamefont{S.}~\bibnamefont{Kumar}},
  \bibinfo{author}{\bibfnamefont{S.~M.} \bibnamefont{Girvin}},
  \bibnamefont{and} \bibinfo{author}{\bibfnamefont{R.~J.}
  \bibnamefont{Schoelkopf}}, \bibinfo{journal}{Nature}
  \textbf{\bibinfo{volume}{431}}, \bibinfo{pages}{162} (\bibinfo{year}{2004}).

\bibitem[{\citenamefont{Regal et~al.}(2008)\citenamefont{Regal, Teufel, and
  Lehnert}}]{Regal2008}
\bibinfo{author}{\bibfnamefont{C.~A.} \bibnamefont{Regal}},
  \bibinfo{author}{\bibfnamefont{J.~D.} \bibnamefont{Teufel}},
  \bibnamefont{and} \bibinfo{author}{\bibfnamefont{K.~W.}
  \bibnamefont{Lehnert}}, \bibinfo{journal}{Nature Phys.}
  \textbf{\bibinfo{volume}{4}}, \bibinfo{pages}{555} (\bibinfo{year}{2008}).

\bibitem[{\citenamefont{Gao et~al.}(2007)\citenamefont{Gao, Zmuidzinas, Mazin,
  Day, and LeDuc}}]{Gao2007}
\bibinfo{author}{\bibfnamefont{J.}~\bibnamefont{Gao}},
  \bibinfo{author}{\bibfnamefont{J.}~\bibnamefont{Zmuidzinas}},
  \bibinfo{author}{\bibfnamefont{B.~A.} \bibnamefont{Mazin}},
  \bibinfo{author}{\bibfnamefont{P.~K.} \bibnamefont{Day}}, \bibnamefont{and}
  \bibinfo{author}{\bibfnamefont{H.~G.} \bibnamefont{LeDuc}},
  \bibinfo{journal}{Appl. Phys. Lett.} \textbf{\bibinfo{volume}{90}},
  \bibinfo{pages}{817} (\bibinfo{year}{2007}).

\bibitem[{\citenamefont{Gao et~al.}(2008{\natexlab{a}})\citenamefont{Gao, Daal,
  Martinis, Vayonakis, Zmuidzinas, Sadoulet, Mazin, Day, and Leduc}}]{Gao2008b}
\bibinfo{author}{\bibfnamefont{J.}~\bibnamefont{Gao}},
  \bibinfo{author}{\bibfnamefont{M.}~\bibnamefont{Daal}},
  \bibinfo{author}{\bibfnamefont{J.~M.} \bibnamefont{Martinis}},
  \bibinfo{author}{\bibfnamefont{A.}~\bibnamefont{Vayonakis}},
  \bibinfo{author}{\bibfnamefont{J.}~\bibnamefont{Zmuidzinas}},
  \bibinfo{author}{\bibfnamefont{B.}~\bibnamefont{Sadoulet}},
  \bibinfo{author}{\bibfnamefont{B.~A.} \bibnamefont{Mazin}},
  \bibinfo{author}{\bibfnamefont{P.~K.} \bibnamefont{Day}}, \bibnamefont{and}
  \bibinfo{author}{\bibfnamefont{H.~G.} \bibnamefont{Leduc}},
  \bibinfo{journal}{Appl. Phys. Lett.} \textbf{\bibinfo{volume}{92}},
  \bibinfo{eid}{212504} (\bibinfo{year}{2008}{\natexlab{a}}).

\bibitem[{\citenamefont{M\"{u}ck et~al.}(2003)\citenamefont{M\"{u}ck, Welzel,
  and Clarke}}]{Michael2003}
\bibinfo{author}{\bibfnamefont{M.}~\bibnamefont{M\"{u}ck}},
  \bibinfo{author}{\bibfnamefont{C.}~\bibnamefont{Welzel}}, \bibnamefont{and}
  \bibinfo{author}{\bibfnamefont{J.}~\bibnamefont{Clarke}},
  \bibinfo{journal}{Appl. Phys. Lett.} \textbf{\bibinfo{volume}{82}},
  \bibinfo{pages}{3266} (\bibinfo{year}{2003}).

\bibitem[{\citenamefont{Spietz et~al.}(2009)\citenamefont{Spietz, Irwin, and
  Aumentado}}]{Spietz2009}
\bibinfo{author}{\bibfnamefont{L.}~\bibnamefont{Spietz}},
  \bibinfo{author}{\bibfnamefont{K.}~\bibnamefont{Irwin}}, \bibnamefont{and}
  \bibinfo{author}{\bibfnamefont{J.}~\bibnamefont{Aumentado}},
  \bibinfo{journal}{Appl. Phys. Lett.} \textbf{\bibinfo{volume}{95}},
  \bibinfo{eid}{092505} (\bibinfo{year}{2009}).

\bibitem[{\citenamefont{Yurke et~al.}(1989)\citenamefont{Yurke, Corruccini,
  Kaminsky, Rupp, Smith, Silver, Simon, and Whittaker}}]{Yurke1989}
\bibinfo{author}{\bibfnamefont{B.}~\bibnamefont{Yurke}},
  \bibinfo{author}{\bibfnamefont{L.~R.} \bibnamefont{Corruccini}},
  \bibinfo{author}{\bibfnamefont{P.~G.} \bibnamefont{Kaminsky}},
  \bibinfo{author}{\bibfnamefont{L.~W.} \bibnamefont{Rupp}},
  \bibinfo{author}{\bibfnamefont{A.~D.} \bibnamefont{Smith}},
  \bibinfo{author}{\bibfnamefont{A.~H.} \bibnamefont{Silver}},
  \bibinfo{author}{\bibfnamefont{R.~W.} \bibnamefont{Simon}}, \bibnamefont{and}
  \bibinfo{author}{\bibfnamefont{E.~A.} \bibnamefont{Whittaker}},
  \bibinfo{journal}{Phys. Rev. A} \textbf{\bibinfo{volume}{39}},
  \bibinfo{pages}{2519} (\bibinfo{year}{1989}).

\bibitem[{\citenamefont{{Castellanos-Beltran}
  et~al.}(2008)\citenamefont{{Castellanos-Beltran}, Irwin, Hilton, Vale, and
  Lehnert}}]{Castellanos-Beltran2008}
\bibinfo{author}{\bibfnamefont{M.~A.} \bibnamefont{{Castellanos-Beltran}}},
  \bibinfo{author}{\bibfnamefont{K.~D.} \bibnamefont{Irwin}},
  \bibinfo{author}{\bibfnamefont{G.~C.} \bibnamefont{Hilton}},
  \bibinfo{author}{\bibfnamefont{L.~R.} \bibnamefont{Vale}}, \bibnamefont{and}
  \bibinfo{author}{\bibfnamefont{K.~W.} \bibnamefont{Lehnert}},
  \bibinfo{journal}{Nature Phys.} \textbf{\bibinfo{volume}{4}},
  \bibinfo{pages}{929} (\bibinfo{year}{2008}).

\bibitem[{\citenamefont{Siddiqi et~al.}(2004)\citenamefont{Siddiqi, Vijay,
  Pierre, Wilson, Metcalfe, Rigetti, Frunzio, and Devoret}}]{Siddiqi2004}
\bibinfo{author}{\bibfnamefont{I.}~\bibnamefont{Siddiqi}},
  \bibinfo{author}{\bibfnamefont{R.}~\bibnamefont{Vijay}},
  \bibinfo{author}{\bibfnamefont{F.}~\bibnamefont{Pierre}},
  \bibinfo{author}{\bibfnamefont{C.~M.} \bibnamefont{Wilson}},
  \bibinfo{author}{\bibfnamefont{M.}~\bibnamefont{Metcalfe}},
  \bibinfo{author}{\bibfnamefont{C.}~\bibnamefont{Rigetti}},
  \bibinfo{author}{\bibfnamefont{L.}~\bibnamefont{Frunzio}}, \bibnamefont{and}
  \bibinfo{author}{\bibfnamefont{M.~H.} \bibnamefont{Devoret}},
  \bibinfo{journal}{Phys. Rev. Lett.} \textbf{\bibinfo{volume}{93}},
  \bibinfo{pages}{207002} (\bibinfo{year}{2004}).

\bibitem[{cal()}]{cal}
\bibinfo{note}{$P_\mathrm{c}$ depends on the critical current of the junction
  and $Q_\mathrm{JPA}$, which are unchanged for the two cases at the same
  temperature of 4K. We measured $P_\mathrm{c, 4K} = -93.2$~dBm using a VNA in
  the calibrated LHe probe test. $A_\mathrm{cal}$ is generated by the same VNA
  at 60.0~dB below $P_\mathrm{c, 4K}$, therefore, $P_\mathrm{cal} =
  -153.2~$dBm. We use this calibration method due to the lack of cooling power
  for a hot-cold load measurement in our ADR . The details of this calibration
  procedure will be published in a separate paper.}

\bibitem[{\citenamefont{Caves}(1982)}]{Caves1982}
\bibinfo{author}{\bibfnamefont{C.~M.} \bibnamefont{Caves}},
  \bibinfo{journal}{Phys. Rev. D} \textbf{\bibinfo{volume}{26}},
  \bibinfo{pages}{1817} (\bibinfo{year}{1982}).

\bibitem[{gai()}]{gain}
\bibinfo{note}{We have simply chosen a 2.3~MHz signal detuning to exclude the
  pump tone from the 1~MHz noise sampling bandwidth and all our noise plots.
  Accordingly, we have used a lower JPA gain G=16 which produces a 5~MHz
  bandwidth (single-sided) to accommodate the signal. It is possible to improve
  the system noise by carefully selecting a smaller signal detuning and a
  higher JPA gain.}

\bibitem[{\citenamefont{Kumar et~al.}(2008)\citenamefont{Kumar, Gao,
  Zmuidzinas, Mazin, LeDuc, and Day}}]{Kumar2008}
\bibinfo{author}{\bibfnamefont{S.}~\bibnamefont{Kumar}},
  \bibinfo{author}{\bibfnamefont{J.}~\bibnamefont{Gao}},
  \bibinfo{author}{\bibfnamefont{J.}~\bibnamefont{Zmuidzinas}},
  \bibinfo{author}{\bibfnamefont{B.~A.} \bibnamefont{Mazin}},
  \bibinfo{author}{\bibfnamefont{H.~G.} \bibnamefont{LeDuc}}, \bibnamefont{and}
  \bibinfo{author}{\bibfnamefont{P.~K.} \bibnamefont{Day}},
  \bibinfo{journal}{Appl. Phys. Lett.} \textbf{\bibinfo{volume}{92}},
  \bibinfo{eid}{123503} (\bibinfo{year}{2008}).

\bibitem[{\citenamefont{Noroozian et~al.}(2009)\citenamefont{Noroozian, Gao,
  Zmuidzinas, LeDuc, and Mazin}}]{Omid2009}
\bibinfo{author}{\bibfnamefont{O.}~\bibnamefont{Noroozian}},
  \bibinfo{author}{\bibfnamefont{J.}~\bibnamefont{Gao}},
  \bibinfo{author}{\bibfnamefont{J.}~\bibnamefont{Zmuidzinas}},
  \bibinfo{author}{\bibfnamefont{H.~G.} \bibnamefont{LeDuc}}, \bibnamefont{and}
  \bibinfo{author}{\bibfnamefont{B.~A.} \bibnamefont{Mazin}}
  (\bibinfo{publisher}{AIP}, \bibinfo{year}{2009}), vol.
  \bibinfo{volume}{1185}, pp. \bibinfo{pages}{148--151}.

\bibitem[{\citenamefont{Gao et~al.}(2008{\natexlab{b}})\citenamefont{Gao, Daal,
  Vayonakis, Kumar, Zmuidzinas, Sadoulet, Mazin, Day, and Leduc}}]{Gao2008a}
\bibinfo{author}{\bibfnamefont{J.}~\bibnamefont{Gao}},
  \bibinfo{author}{\bibfnamefont{M.}~\bibnamefont{Daal}},
  \bibinfo{author}{\bibfnamefont{A.}~\bibnamefont{Vayonakis}},
  \bibinfo{author}{\bibfnamefont{S.}~\bibnamefont{Kumar}},
  \bibinfo{author}{\bibfnamefont{J.}~\bibnamefont{Zmuidzinas}},
  \bibinfo{author}{\bibfnamefont{B.}~\bibnamefont{Sadoulet}},
  \bibinfo{author}{\bibfnamefont{B.~A.} \bibnamefont{Mazin}},
  \bibinfo{author}{\bibfnamefont{P.~K.} \bibnamefont{Day}}, \bibnamefont{and}
  \bibinfo{author}{\bibfnamefont{H.~G.} \bibnamefont{Leduc}},
  \bibinfo{journal}{Appl. Phys. Lett.} \textbf{\bibinfo{volume}{92}},
  \bibinfo{pages}{152505} (\bibinfo{year}{2008}{\natexlab{b}}).

\bibitem[{\citenamefont{Phillips}(1972)}]{Phillips1972}
\bibinfo{author}{\bibfnamefont{W.~A.} \bibnamefont{Phillips}},
  \bibinfo{journal}{J. Low Temp.Phys.} \textbf{\bibinfo{volume}{7}},
  \bibinfo{pages}{351} (\bibinfo{year}{1972}).

\bibitem[{\citenamefont{Martinis et~al.}(2005)\citenamefont{Martinis, Cooper,
  McDermott, Steffen, Ansmann, Osborn, Cicak, Oh, Pappas, Simmonds
  et~al.}}]{Martinis2005}
\bibinfo{author}{\bibfnamefont{J.~M.} \bibnamefont{Martinis}},
  \bibinfo{author}{\bibfnamefont{K.~B.} \bibnamefont{Cooper}},
  \bibinfo{author}{\bibfnamefont{R.}~\bibnamefont{McDermott}},
  \bibinfo{author}{\bibfnamefont{M.}~\bibnamefont{Steffen}},
  \bibinfo{author}{\bibfnamefont{M.}~\bibnamefont{Ansmann}},
  \bibinfo{author}{\bibfnamefont{K.~D.} \bibnamefont{Osborn}},
  \bibinfo{author}{\bibfnamefont{K.}~\bibnamefont{Cicak}},
  \bibinfo{author}{\bibfnamefont{S.}~\bibnamefont{Oh}},
  \bibinfo{author}{\bibfnamefont{D.~P.} \bibnamefont{Pappas}},
  \bibinfo{author}{\bibfnamefont{R.~W.} \bibnamefont{Simmonds}},
  \bibnamefont{et~al.}, \bibinfo{journal}{Phys. Rev. Lett.}
  \textbf{\bibinfo{volume}{95}}, \bibinfo{pages}{210503}
  (\bibinfo{year}{2005}).

\bibitem[{\citenamefont{Yu}(2004)}]{Yu2004}
\bibinfo{author}{\bibfnamefont{C.~C.} \bibnamefont{Yu}}, \bibinfo{journal}{J.
  Low Temp.Phys.} \textbf{\bibinfo{volume}{137}}, \bibinfo{pages}{251}
  (\bibinfo{year}{2004}).

\bibitem[{\citenamefont{Constantin et~al.}(2009)\citenamefont{Constantin, Yu,
  and Martinis}}]{Constantin2009}
\bibinfo{author}{\bibfnamefont{M.}~\bibnamefont{Constantin}},
  \bibinfo{author}{\bibfnamefont{C.~C.} \bibnamefont{Yu}}, \bibnamefont{and}
  \bibinfo{author}{\bibfnamefont{J.~M.} \bibnamefont{Martinis}},
  \bibinfo{journal}{Phys. Rev. B} \textbf{\bibinfo{volume}{79}},
  \bibinfo{pages}{094520} (\bibinfo{year}{2009}).

\end{thebibliography}

%
\newpage
\newpage
\pagebreak
\pagebreak

\end{document}